# Local spectroscopy of a gate-switchable moiré quantum anomalous Hall insulator

Canxun Zhang[1,2,3,6], Tiancong Zhu[1,2,6]*, Tomohiro Soejima[1,6], Salman Kahn[1,2,6], Kenji Watanabe[4], Takashi Taniguchi[5], Alex Zettl[1,2,3], Feng Wang[1,2,3], Michael P. Zaletel[1,2]*, Michael F. Crommie[1,2,3]*

[1]Department of Physics, University of California, Berkeley, CA 94720, USA.

[2]Materials Sciences Division, Lawrence Berkeley National Laboratory, Berkeley, CA 94720, USA.

[3]Kavli Energy NanoScience Institute at the University of California, Berkeley and the Lawrence Berkeley National Laboratory, Berkeley, CA 94720, USA.

[4]Research Center for Electronic and Optical Materials, National Institute for Materials Science, 1-1 Namiki, Tsukuba 305-0044, Japan.

[5]Research Center for Materials Nanoarchitectonics, National Institute for Materials Science, 1-1 Namiki, Tsukuba 305-0044, Japan.

[6]These authors contributed equally: Canxun Zhang, Tiancong Zhu, Tomohiro Soejima, Salman Kahn.

*Email: tiancongzhu@berkeley.edu; mikezaletel@berkeley.edu; crommie@berkeley.edu.

## Abstract

In recent years, correlated insulating states, unconventional superconductivity, and topologically non-trivial phases have all been observed in several moiré heterostructures. However, understanding of the physical mechanisms behind these phenomena is hampered by the lack of local electronic structure data. Here, we use scanning tunnelling microscopy and spectroscopy to demonstrate how the interplay between correlation, topology, and local atomic structure determines the behaviour of electron-doped twisted monolayer-bilayer graphene. Through gate- and magnetic field-dependent measurements, we observe local spectroscopic signatures indicating a quantum anomalous Hall insulating state with a total Chern number of ±2 at a doping level of three electrons per moiré unit cell. We show that the sign of the Chern number and associated magnetism can be electrostatically switched only over a limited range of twist angle and sample hetero-strain values. This results from a competition between the orbital magnetization of filled bulk bands and chiral edge states, which is sensitive to strain-induced distortions in the moiré superlattice.

## Introduction

Van der Waals stacking of twisted two-dimensional (2D) atomic sheets provides a versatile platform for engineering exotic electronic states through rotational misalignment that folds dispersive electronic bands into flat mini-bands within a moiré Brillouin zone.[1,2] The resulting suppression of kinetic energy relative to electron-electron interactions can lead to correlated insulating states as well as unconventional superconductivity.[3,4] Moiré flat bands also inherit the large Berry curvature of the individual atomic layers which can result in topologically non-trivial phases.[5-7] Electron-doped twisted monolayer-bilayer graphene (tMBLG)—a graphene monolayer rotationally misaligned with a Bernal-stacked bilayer—stands out among these since it exhibits the quantum anomalous Hall (QAH) effect (i.e., quantized Hall conductance in the absence of external magnetic field) accompanied by doping-controlled switching of its Chern number, an effect not observed in other moiré QAH systems.[8] Such behaviour is expected to be sensitive to local structural parameters such as twist angle and hetero-strain (i.e., the relative strain between adjacent layers). For example, twist angle directly affects the moiré mini-band structure while even small hetero-strains (< 0.5%) can be magnified by the moiré superlattice to induce large moiré distortions, thus altering the energetics of mini-bands and the behaviour of emergent correlated and topological phases.[9,10] Understanding the rich physics of moiré systems requires understanding the relationship between exotic electronic phases and local structure, something difficult to achieve using macroscopic probes that only explore spatially-averaged behaviour.

Here we show how scanning tunnelling microscopy and spectroscopy (STM/STS) enables determination of how changes in local structure alter correlated and topological electronic behaviour in tMBLG field-effect transistor devices. We find that tuning the electron doping concentration of tMBLG results in the emergence of charge gaps observable to STS at filling levels $v = 2$ and $v = 3$ (i.e., two and three electrons per moiré unit cell), indicating the formation of correlated insulating states. STS performed in an out-of-plane magnetic field allows us to detect non-trivial topology in the $v = 3$ QAH insulating state which has total Chern number $C_{\text{tot}} = \pm 2$, and to demonstrate its dependence on local twist angle and hetero-strain. In addition to observing strong variation of correlation and topological properties at different twist angles, we find that regions having nearly identical twist angle but different hetero-strain values exhibit very different behaviour. In the small-strain regime, the correlation gap evolves into two separate gaps at different gate voltages that correspond to $C_{\text{tot}} = +2$ and $C_{\text{tot}} = -2$, indicating doping-controlled switching of valley polarization consistent with previous electrical transport results.[8] Such behaviour is absent, however, when large hetero-strain is present, in which case only a single correlation gap with $C_{\text{tot}} = +2$ is observed. This behaviour can be understood using a continuum model for tMBLG that reveals how Chern number switching results from a competition between the bulk and edge contributions to orbital magnetization that is highly sensitive to local hetero-strain. These results demonstrate the crucial role that local structural parameters play in shaping correlation and topological effects in twisted moiré systems.

**Results**

**Correlated insulating behaviour at integer fillings.** Figure 1a shows a schematic of our experiment, which incorporates a gate-tunable graphene device into an STM measurement geometry. A Bernal-stacked bilayer graphene is placed on top of a graphene monolayer with a twist angle $\theta$ between them, and the stack is supported by a hexagonal boron nitride (hBN) substrate placed on a $SiO_2$/Si wafer (Methods, Supplementary Fig. 1). The carrier density $n$ of the graphene stack can be tuned continuously via voltage $V_G$ applied to the Si back-gate. Our devices were annealed in ultra-high vacuum before being loaded into the STM system at $T = 4.6$ K for measurement (Methods). Figure 1b shows a representative topographic image of the monolayer-bilayer moiré pattern which exhibits an average wavelength of $l_M = 11.2$ nm, from which we extracted a local twist angle of $\theta = 1.25°$ (Methods). Within each moiré unit cell (dashed box) we observe three representative regions with different apparent heights that correspond to the three local tMBLG stacking orders: BAB, ABC, and AAB (Supplementary Note 1).

We access correlated electronic states of tMBLG by tuning the carrier concentration via $V_G$ and performing d$I$/d$V$ spectroscopy. Figure 1c shows a density plot of gate-dependent d$I$/d$V$ spectra obtained in the BAB region. Estimation of the device capacitance allows us to convert $V_G$ to the filling factor $v$, defined as the average number of electrons/holes per moiré unit cell referenced to charge neutrality (Methods). At $v = 0$ ($V_G = 0$ V) we observe two narrow peaks in the d$I$/d$V$ spectrum that are centred at $V_{Bias} = 6$ mV and $V_{Bias} = -14$ mV (Fig. 1d) that we identify as originating from van Hove singularities of the four-fold degenerate conduction flat band (CFB) and valence flat band (VFB). Increasing $V_G$ leads to partial occupation of the CFB and shifts both peaks toward lower energy. As the filling level approaches $v = 2$ the CFB peak gradually splits into two branches, CFB– and CFB+, that are located below and above the Fermi energy $E_F$ ($V_{Bias} = 0$ mV). At $v = 2$ ($V_G = 31.5$ V) these two branches have roughly the same spectral weight and a clear charge gap can be observed across $E_F$ (Fig. 1e). As the doping level is further increased from $v = 2$ to $v = 2.5$ ($V_G = 39$ V), the energy splitting between CFB– and CFB+ becomes smaller and the gap feature evolves into a shallow dip (Fig. 1f). At $v = 3$ ($V_G = 47$ V) an insulating gap reappears at $E_F$ with CFB– having significantly greater weight compared to CFB+ (Fig. 1g). Finally, at $v = 4$ ($V_G = 62.5$ V, full filling of the CFB) the CFB– and CFB+ branches merge into a single peak that lies completely below $E_F$ (Fig. 1h). The presence of charge gaps at $v = 2$ and $v = 3$ demonstrates the formation of correlated insulating states at these filling factors, corroborating results from previous electrical transport studies[8,11-13] (Supplementary Note 2, Supplementary Figs. 2, 3).

**Gate-switchable QAH insulating state.** To discern the nature of the $v = 2, 3$ correlated insulating states in tMBLG, we applied an out-of-plane magnetic field **B** = (0, 0, $B$) to our sample and performed gate-dependent d$I$/d$V$ spectroscopy. Figure 2a-c shows density plots of gate-dependent d$I$/d$V$ spectra measured near $v = 2$ for $B = 0, 1, 2$ T, respectively. The insulating gap feature, marked by vanishing d$I$/d$V$ at $E_F$ and maximum CFB peak splitting, always appears at the same filling level (white arrows) regardless of the $B$ value. d$I$/d$V$ spectra measured near $v = 3$ (Fig. 2d-i), however, exhibit very different field-dependent behaviour. The charge gap (white arrows) is seen to remain constant in energy splitting (Supplementary Note 3) but to evolve into two separate gaps for $B > 0$ T. These two gaps bracket $v = 3$ and split away from it as $B$ increases.

We can better visualize the magnetic field evolution of the $v = 3$ correlated insulating state by plotting normalized d$I$/d$V$ at $V_{Bias} = 0$ mV ($E_F$) as a function of both $v$ ($V_G$) and $B$ (Fig. 2j). The dark region in the plot indicates vanishing d$I$/d$V$ due to the emergence of a charge gap, which forms a V-shape (white dashed lines) that is roughly symmetric about the $v = 3$ horizontal line. This linear scaling of the correlation gap position with magnetic field is reminiscent of correlated Chern insulating states reported previously in magic-angle twisted bilayer graphene (MA-tBLG)[14-16] where the change in carrier concentration $n$ of a Chern insulating state is related to the out-of-plane field $B$ through the Středa formula $\frac{\Delta n}{\Delta B} = \frac{C_{tot}}{\Phi_0}$ ($C_{tot}$ is the total Chern number and $\Phi_0 = h/e$ is the magnetic flux quantum).[17] Our observations thus imply that the $v = 3$ insulating state in tMBLG has Chern number $C_{tot} = \pm 2$ as derived from the slope of the lines in Fig. 2j. Two significant differences, however, distinguish our results from those reported in MA-tBLG. First, $\Delta n/\Delta B$ linear scaling is observed in MA-tBLG only under high external fields ($B > 3$ T) that break time-reversal symmetry and stabilize the Chern insulating states, whereas such behaviour in tMBLG can be resolved in fields as low as $B = 0.2$ T (Supplementary Fig. 4) and can be traced back to $B = 0$ T. Combined with the robust charge gap at $v = 3$, this indicates that the zero-field ground state of tMBLG is a topologically non-trivial QAH insulator with spontaneous time-reversal symmetry breaking (Supplementary Note 4). Second, each non-zero integer filling of MA-tBLG features only a single correlation gap that shifts monotonically with increasing $B$, while in tMBLG we observe two separate gaps corresponding to $C_{tot} = +2$ and $C_{tot} = -2$. This indicates that the total Chern number for the tMBLG QAH state is switchable between +2 and –2 by simply tuning the carrier concentration across $v = 3$.[8]

**Tuning Chern number switching with twist angle and strain.** Simultaneous structural measurement via STM and local electronic characterization via STS provide a unique opportunity to investigate how correlation and topological effects in tMBLG are affected by local structural variations at the moiré scale. Figure 3a summarizes our experimental results as a function of local twist angle and local hetero-strain obtained through analysis of moiré anisotropy in our STM topographs (Methods). While the emergence of an insulating gap at $v = 2$ is robust in all of our data, the behaviour at $v = 3$ depends strongly on both local twist angle and local hetero-strain. Gate-tunability of the Chern number is only observed when the twist angle is between 1.25° and 1.28°, as indicated by the green data points in Fig. 3a. When the twist angle increases slightly above this range (orange data points) the charge gap at $v = 3$ persists but Chern number switching is suppressed. Fig. 3b-d shows representative data from this regime in which the gap feature (white arrows) evolves monotonically toward higher filling factors as the magnetic field is increased instead of developing into two separate gaps. When the twist angle deviates even further the correlation gap at $v = 3$ disappears (red data points in Fig. 3a; see Supplementary Fig. 5).

To reveal the effect of hetero-strain, we directly compare two regions with almost identical twist angle (~1.26°) but different hetero-strain values (0.10% versus 0.24%) for the same device, thus allowing other variables such as carrier concentration, electric field, and correlation strength to be kept mostly constant. In the region with a smaller hetero-strain (Fig. 3e), the $v = 3$ insulating gap develops into two branches under application of an out-of-plane magnetic field (Fig. 3f), indicating gate-induced switching between $C_{tot} = +2$ and $C_{tot} = -2$ QAH insulating states. In contrast, the region with a larger hetero-strain (Fig. 3g) exhibits only one branch of the $v = 3$ insulating gap (Fig. 3h; see Supplementary Fig. 6), corresponding to $C_{tot} = +2$ with no gate-controlled switching.

**Discussion**

The emergence of correlated QAH insulating states in electron-doped tMBLG can be understood as resulting from spontaneous symmetry breaking driven by electron-electron Coulomb interactions. The CFB of tMBLG is four-fold degenerate due to spin and valley degrees of freedom. Each CFB sub-band in the graphene K+ (K–) valley hosts a non-zero Chern number of $C = +2$ (–2) due to the large Berry curvature inherited from constituent graphene layers.[18,19] At integer fillings (e.g., $v = 2$ and $v = 3$) strong correlation can drive spontaneous polarization along a certain axis in the spin-valley space, splitting the CFB into occupied lower sub-bands (CFB–) and unoccupied upper sub-bands (CFB+) separated by a charge gap as observed in the experimental d$I$/d$V$ (Fig. 1e,g). At $v = 3$ the spin-valley polarization leads to breaking of time-reversal symmetry and topologically non-trivial states with $C_{tot} \neq 0$. Figure 4a shows one possible filling configuration with double occupancy of CFB sub-bands in the K+ valley, resulting in a QAH insulating state with $C_{tot} = +2$. Similarly, double occupancy of K– valley sub-bands can result in a $C_{tot} = -2$ state that is energetically equivalent to the $C_{tot} = +2$ state in the absence of an external magnetic field.

The large Berry curvature in the tMBLG moiré flat bands produces out-of-plane orbital magnetic moments that respond to an external magnetic field. The competition between bulk orbital magnetization ($M_{bulk}$) due to self-rotation of electron wave packets and edge orbital magnetization ($M_{edge}$) due to circulation of electrons in the topologically-protected in-gap chiral edge states is responsible for the gate-controlled switching behaviour at $v = 3$ (spin magnetism plays no explicit role in the switching due to negligible spin-orbit coupling in tMBLG).[8] When the chemical potential resides in the bulk correlation gap the sign of $M_{edge}$ is the same as the sign of $C_{tot}$ and its magnitude is determined by how much the edge state band is filled.[20] The sign and magnitude of $M_{bulk}$, on the other hand, are sensitive to the detailed band structure and are not simply related to $C_{tot}$ other than the fact that a reversal in $C_{tot}$ is accompanied by a reversal in $M_{bulk}$ (Supplementary Note 6). The total orbital magnetization is the sum of $M_{bulk}$ and $M_{edge}$ and can be altered electrostatically by changing the filling of the edge states.

Our data is consistent with the special case illustrated in Fig. 4b,c which shows magnetic states for both $C_{tot} = +2$ and $C_{tot} = -2$ near $v = 3$ under the condition that $M_{bulk}$ and $M_{edge}$ are antiparallel and $|M_{edge}| > |M_{bulk}|$. When $E_F$ is placed at the top of the correlation gap ($v \gtrsim 3$, Fig. 4b), the chiral edge states are fully filled, and the total orbital magnetization is in the same direction as $M_{edge}$. When $E_F$ is moved to the bottom of the correlation gap ($v \lesssim 3$, Fig. 4c), however, the edge states are empty and the total magnetization is in the direction of $M_{bulk}$. In an external magnetic field the system will be driven into a magnetic ground state that aligns the total orbital magnetic moment with the applied field to minimize the orbital Zeeman energy. For the case shown here the energetically favourable states are $C_{tot} = +2$ for $v \gtrsim 3$ and $C_{tot} = -2$ for $v \lesssim 3$, thus illustrating how electrostatic gating can cause Chern number switching in tMBLG consistent with our experimental observations.

We have performed theoretical simulations that support the assumptions made in the switching picture described above (i.e., that $M_{bulk}$ and $M_{edge}$ are antiparallel and that $|M_{edge}| > |M_{bulk}|$) and that also explain why some tMBLG regions do not exhibit switching. For the experimental regimes shown in Fig. 3e-h (where the tMBLG twist angle lies close to 1.26°) we find that the key physical parameter that controls magnetic switching functionality is hetero-strain. To see this we calculated the strain-induced behaviour of $M_{bulk}$ and $M_{edge}$ for a continuum model of 1.26° tMBLG in the $v = 3$ QAH insulating state with $E_F$ set to the top of the correlation gap (a single phenomenological parameter characterizing the electron-electron interaction strength is included similar to Ref. [8]; see Supplementary Note 6 for details). Fig. 4d shows a plot of the resulting $M_{edge}$ and $M_{bulk}$ for $C_{tot} = +2$ (top) and $C_{tot} = -2$ (bottom) as a function of hetero-strain with all other parameters kept constant. For small strain (< 0.15%) $|M_{edge}|$ is significantly greater than $|M_{bulk}|$ while $M_{edge}$ and $M_{bulk}$ are anti-aligned. For large hetero-strain (> 0.15%), however, the situation changes significantly. The sign of $M_{edge}$ remains the same as $C_{tot}$, but $M_{bulk}$ flips its sign due to a strain-induced redistribution of Berry curvature and band dispersion throughout the mini-Brillouin zone. This divides the strain parameter space into two regimes separated by the vertical dashed line in Fig.4d: the small-strain regime where $M_{bulk}$ and $M_{edge}$ have opposite sign (and switching can occur), and the large-strain regime where they have the same sign (and switching does not occur). To better illustrate this behaviour, Fig. 4e shows a schematic of the resulting orbital Zeeman energy for $B > 0$ as a function of filling factor near $\nu = 3$. Here the solid lines show the energetically favourable ground state and the dashed lines show the unfavourable state having opposite $C_{tot}$. For small hetero-strain $M_{bulk}$ dominates for $\nu \lesssim 3$ (when the edge state band is empty) and results in $C_{tot} = -2$ whereas $M_{edge}$ dominates for $\nu \gtrsim 3$ (as the edge states are filled), resulting in a switching to $C_{tot} = +2$. For large hetero-strain, $M_{bulk}$ and $M_{edge}$ both line up parallel to the applied field at all fillings and the system stays at $C_{tot} = +2$ without switching, consistent with our experimental observations.

In conclusion, we have observed local spectroscopic signatures of strong correlation and non-trivial topology at the moiré scale in tMBLG, and have demonstrated local control of the QAH Chern number via electrostatic gating. Combining STM and STS allows us to characterize the topological and magnetic switching phase diagram of tMBLG in the parameter space of local twist angle and hetero-strain. We observe magnetic switching only at low strain, revealing the sensitive interplay between correlation, topology, and local structural parameters that determines electronic and magnetic ground states in twisted moiré systems. This provides insight into the many phase diagrams observed in related moiré systems via measurements that average over regions having different structural parameters, and creates new opportunities for future manipulation of QAH insulating domains and the chiral edge states that lie between them.

**Methods**

**Sample preparation.** Samples were prepared using the "flip-chip" method[21] followed by a forming-gas anneal.[22,23] Electrical contacts were made by evaporating Cr/Au (5 nm/70 nm) through a silicon nitride shadow-mask onto the heterostructure. The sample surface cleanliness was confirmed using contact-AFM prior to STM measurements. Samples were annealed at 300 °C overnight in ultra-high vacuum before insertion into the low-temperature STM stage.

**STM/STS measurements.** STM/STS measurements were performed in a commercial CreaTec LT-STM held at $T$ = 4.6 K using tungsten (W) tips. STM tips were prepared on a Cu(111) surface and calibrated against the Cu(111) Shockley surface state before measurements to avoid tip artifacts. A voltage $V_G$ was applied to the Si back-gate to change the carrier density $n$ in the tMBLG stack ($n = \frac{\varepsilon_D \varepsilon_0 V_G}{e d_D}$ where $\varepsilon_D \approx 3.6$ is the average out-of-plane dielectric constant of hBN and SiO$_2$, $\varepsilon_0$ is the vacuum permittivity, $e$ is the elementary charge, and $d_D$ = 335 nm is the thickness of the dielectric layers). d$I$/d$V$ spectra were recorded using standard lock-in techniques with a small bias modulation $V_{RMS}$ = 1 mV at 613 Hz. All STM images were edited using WSxM software.[24]

**Determination of local twist angle and hetero-strain.** The local twist angle and hetero-strain were determined by analysing the monolayer-bilayer moiré pattern. The moiré wavelength was obtained for three directions by measuring the spatial separation between peaks in STM topographs and averaging over several moiré unit cells. The reciprocal primitive vectors $K_i$ ($i$ = 0, 1, 2) were derived through Fourier transform analysis. In the presence of hetero-strain, $K_i$ can be approximately written as $K_i = k\left(\theta + (1+v_P)\epsilon \cos\left(\alpha + i\frac{2\pi}{3}\right)\sin\left(\alpha + i\frac{2\pi}{3}\right)\right)$ where $k$ = 4.694 nm$^{-1}$ is the length of the graphene reciprocal primitive vectors, $\theta$ is the twist angle, $\epsilon$ is the hetero-strain amplitude, $v_P$ = 0.16 is Poisson's ratio for graphene, and $\alpha$ is the angle between the principal axis of the strain tensor and one of the graphene reciprocal primitive vectors.[25] This allows us to solve both $\theta$ and $\epsilon$ from the extracted $K_i$.

**Data Availability**

The data that support the plots within this paper and the findings of this study are provided in the Source Data file.

**Code Availability**

The computer codes that support the plots within this paper and the findings of this study are available from the corresponding authors upon request.

**References**

1   Trambly de Laissardière, G., Mayou, D. & Magaud, L. Localization of Dirac Electrons in Rotated Graphene Bilayers. *Nano Letters* **10**, 804-808 (2010). https://doi.org:10.1021/nl902948m
2   Bistritzer, R. & MacDonald, A. H. Moiré bands in twisted double-layer graphene.

*Proceedings of the National Academy of Sciences* **108**, 12233-12237 (2011). https://doi.org:10.1073/pnas.1108174108

3   Cao, Y. *et al.* Correlated insulator behaviour at half-filling in magic-angle graphene superlattices. *Nature* **556**, 80-84 (2018). https://doi.org:10.1038/nature26154

4   Cao, Y. *et al.* Unconventional superconductivity in magic-angle graphene superlattices. *Nature* **556**, 43-50 (2018). https://doi.org:10.1038/nature26160

5   Serlin, M. *et al.* Intrinsic quantized anomalous Hall effect in a moiré heterostructure. *Science* **367**, 900-903 (2020). https://doi.org:10.1126/science.aay5533

6   Sharpe, A. L. *et al.* Emergent ferromagnetism near three-quarters filling in twisted bilayer graphene. *Science* **365**, 605-608 (2019). https://doi.org:10.1126/science.aaw3780

7   Lu, X. *et al.* Superconductors, orbital magnets and correlated states in magic-angle bilayer graphene. *Nature* **574**, 653-657 (2019). https://doi.org:10.1038/s41586-019-1695-0

8   Polshyn, H. *et al.* Electrical switching of magnetic order in an orbital Chern insulator. *Nature* **588**, 66-70 (2020). https://doi.org:10.1038/s41586-020-2963-8

9   Bi, Z., Yuan, N. F. Q. & Fu, L. Designing flat bands by strain. *Physical Review B* **100**, 035448 (2019). https://doi.org:10.1103/PhysRevB.100.035448

10  Parker, D. E., Soejima, T., Hauschild, J., Zaletel, M. P. & Bultinck, N. Strain-Induced Quantum Phase Transitions in Magic-Angle Graphene. *Physical Review Letters* **127**, 027601 (2021). https://doi.org:10.1103/PhysRevLett.127.027601

11  Chen, S. *et al.* Electrically tunable correlated and topological states in twisted monolayer–bilayer graphene. *Nature Physics* **17**, 374-380 (2021). https://doi.org:10.1038/s41567-020-01062-6

12  Xu, S. *et al.* Tunable van Hove singularities and correlated states in twisted monolayer–bilayer graphene. *Nature Physics* **17**, 619-626 (2021). https://doi.org:10.1038/s41567-021-01172-9

13  He, M. *et al.* Competing correlated states and abundant orbital magnetism in twisted monolayer-bilayer graphene. *Nature Communications* **12**, 4727 (2021). https://doi.org:10.1038/s41467-021-25044-1

14  Nuckolls, K. P. *et al.* Strongly correlated Chern insulators in magic-angle twisted bilayer graphene. *Nature* **588**, 610-615 (2020). https://doi.org:10.1038/s41586-020-3028-8

15  Choi, Y. *et al.* Correlation-driven topological phases in magic-angle twisted bilayer graphene. *Nature* **589**, 536-541 (2021). https://doi.org:10.1038/s41586-020-03159-7

16  Das, I. *et al.* Symmetry-broken Chern insulators and Rashba-like Landau-level crossings in magic-angle bilayer graphene. *Nature Physics* **17**, 710-714 (2021). https://doi.org:10.1038/s41567-021-01186-3

17  Středa, P. Quantised Hall effect in a two-dimensional periodic potential. *Journal of Physics C: Solid State Physics* **15**, L1299-L1303 (1982). https://doi.org:10.1088/0022-3719/15/36/006

18  Park, Y., Chittari, B. L. & Jung, J. Gate-tunable topological flat bands in twisted monolayer-bilayer graphene. *Physical Review B* **102**, 035411 (2020). https://doi.org:10.1103/PhysRevB.102.035411

19  Rademaker, L., Protopopov, I. V. & Abanin, D. A. Topological flat bands and correlated states in twisted monolayer-bilayer graphene. *Physical Review Research* **2**, 033150 (2020). https://doi.org:10.1103/PhysRevResearch.2.033150

20  Zhu, J., Su, J.-J. & MacDonald, A. H. Voltage-Controlled Magnetic Reversal in Orbital Chern Insulators. *Physical Review Letters* **125**, 227702 (2020). https://doi.org:10.1103/PhysRevLett.125.227702

21  Cui, X. *et al.* Low-Temperature Ohmic Contact to Monolayer $MoS_2$ by van der Waals Bonded Co/h-BN Electrodes. *Nano Letters* **17**, 4781-4786 (2017). https://doi.org:10.1021/acs.nanolett.7b01536

22  Garcia, A. G. F. *et al.* Effective Cleaning of Hexagonal Boron Nitride for Graphene Devices. *Nano Letters* **12**, 4449-4454 (2012). https://doi.org:10.1021/nl3011726

23  Au - Jung, H. S. *et al.* Fabrication of Gate-tunable Graphene Devices for Scanning Tunneling Microscopy Studies with Coulomb Impurities. *JoVE*, e52711 (2015). https://doi.org:doi:10.3791/52711

24  Horcas, I. *et al.* WSXM: A software for scanning probe microscopy and a tool for nanotechnology. *Review of Scientific Instruments* **78**, 013705 (2007). https://doi.org:10.1063/1.2432410

25  Kerelsky, A. *et al.* Maximized electron interactions at the magic angle in twisted bilayer graphene. *Nature* **572**, 95-100 (2019). https://doi.org:10.1038/s41586-019-1431-9

**Acknowledgements**

The authors thank Birui Yang for technical support. This research was supported by the Center for Novel Pathways to Quantum Coherence in Materials, an Energy Frontier Research Center funded by the US Department of Energy, Office of Science, Basic Energy Sciences (STM/STS measurements and analysis). Support was also provided by the Director, Office of Science, Office of Basic Energy Sciences, Materials Sciences and Engineering Division of the US Department of Energy, under contract number DE-AC02-05CH11231, within the van der Waals Heterostructures program (KCWF16) (device architecture development); by the Molecular Foundry at LBNL, which is funded by the Director, Office of Science, Office of Basic Energy Sciences, Scientific User Facilities Division, of the US Department of Energy under Contract No. DE-AC02-05CH11231 (graphene layer characterization); as well as the National Science Foundation Award DMR-2221750 (device fabrication and characterization). T.S. and M.Z. were funded by the US Department of Energy, Office of Science, Office of Basic Energy Sciences, Materials Sciences and Engineering Division under Contract No. DE-AC02-05-CH11231 (Theory of Materials program KC2301). K.W. and T.T. acknowledge support from JSPS KAKENHI (Grant Numbers 20H00354, 21H05233, and 23H02052) and World Premier International Research Center Initiative (WPI), MEXT, Japan. C.Z. acknowledges support from a Kavli ENSI Philomathia Graduate Student Fellowship. T.S. acknowledges fellowship support from the Masason Foundation.

**Author Contributions**

C.Z., T.Z., S.K., and M.F.C. initiated and conceived the research. C.Z. and T.Z. carried out STM/STS measurements and analyses. M.F.C. supervised STM/STS measurements and analyses. S.K. prepared gate-tunable devices. A.Z. and M.F.C. supervised device preparations. T.T. and K.W. provided the hBN crystals. T.S. performed theoretical calculations and analyses. M.P.Z. supervised theoretical calculations and analyses. C.Z., T.Z., and M.F.C. wrote the manuscript with help from all authors. All authors contributed to the scientific discussion.

**Competing Interests**

The authors declare no competing interests.

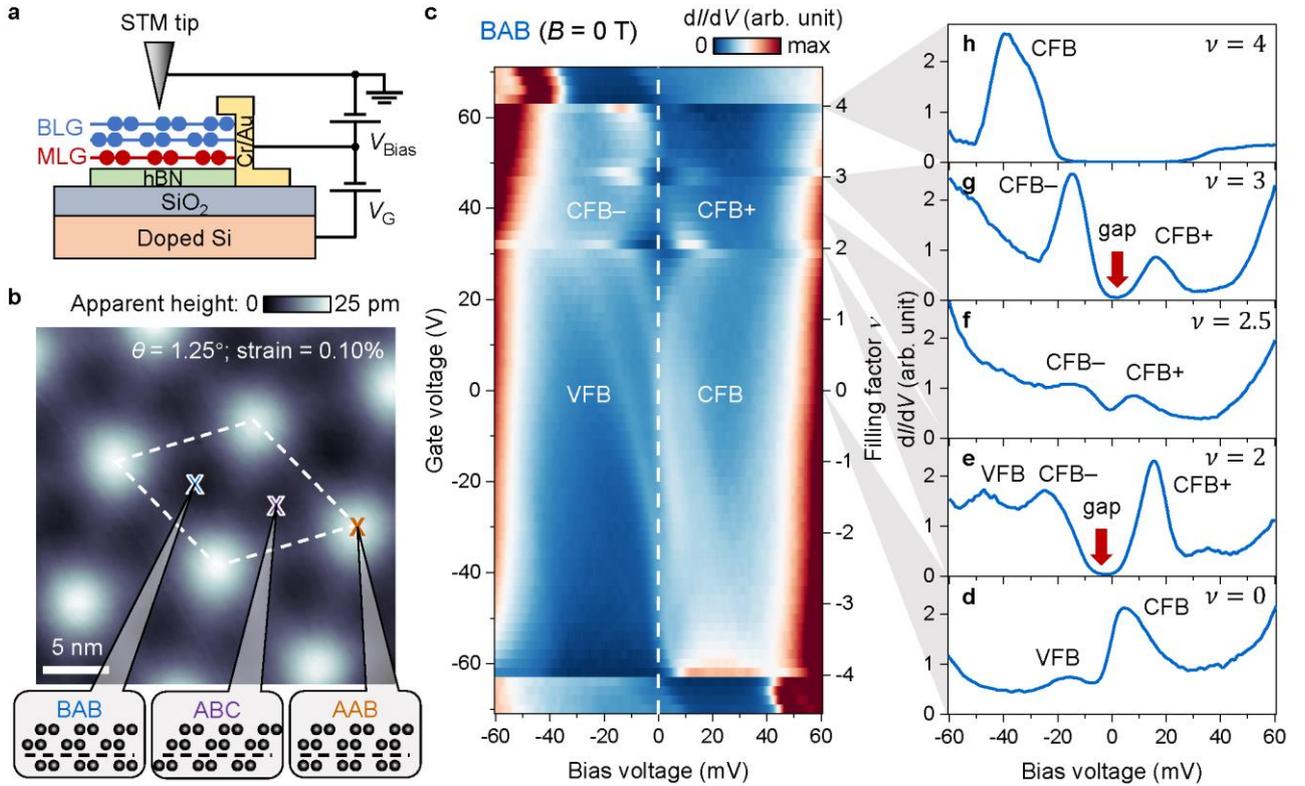

**Figure 1: Correlated insulating states in gate-tunable tMBLG. a**, Schematic of the gate-tunable tMBLG device used in our STM/STS measurements. MLG = monolayer graphene, BLG = Bernal-stacked bilayer graphene. $V_{Bias}$ is the sample bias voltage and $V_G$ is the gate voltage referenced to the sample. **b**, Representative STM topographic image of tMBLG ($V_{Bias} = -1$ V, tunnelling current $I_0 = 0.02$ nA). The dashed box outlines the moiré unit cell. The local stacking orders BAB, ABC, and AAB are shown in the side view. **c**, Gate-dependent d$I$/d$V$ density plot for the BAB stacking region over the gate range $-70$ V $\leq V_G \leq 70$ V. The vertical dashed line denotes the Fermi energy. **d-h**, d$I$/d$V$ spectra measured at (**d**) $V_G = 0$ V ($\nu = 0$), (**e**) $V_G = 31.5$ V ($\nu = 2$), (**f**) $V_G = 39$ V ($\nu = 2.5$), (**g**)

$V_G$ = 47 V ($v$ = 3), and (**h**) $V_G$ = 62.5 V ($v$ = 4). Spectroscopy parameters: modulation voltage $V_{RMS}$ = 1 mV; setpoint $V_{Bias}$ = 100 mV, $I_0$ = 1.15 nA for –70 V ≤ $V_G$ ≤ –2 V in **c**; setpoint $V_{Bias}$ = –100 mV, $I_0$ = 0.8 nA for 0 V ≤ $V_G$ ≤ 70 V in **c** and **d**; setpoint $V_{Bias}$ = –60 mV, $I_0$ = 0.5 nA for **e-h**. VFB = valence flat band, CFB = conduction flat band, CFB– = lower branch of CFB, CFB+ = upper branch of CFB.

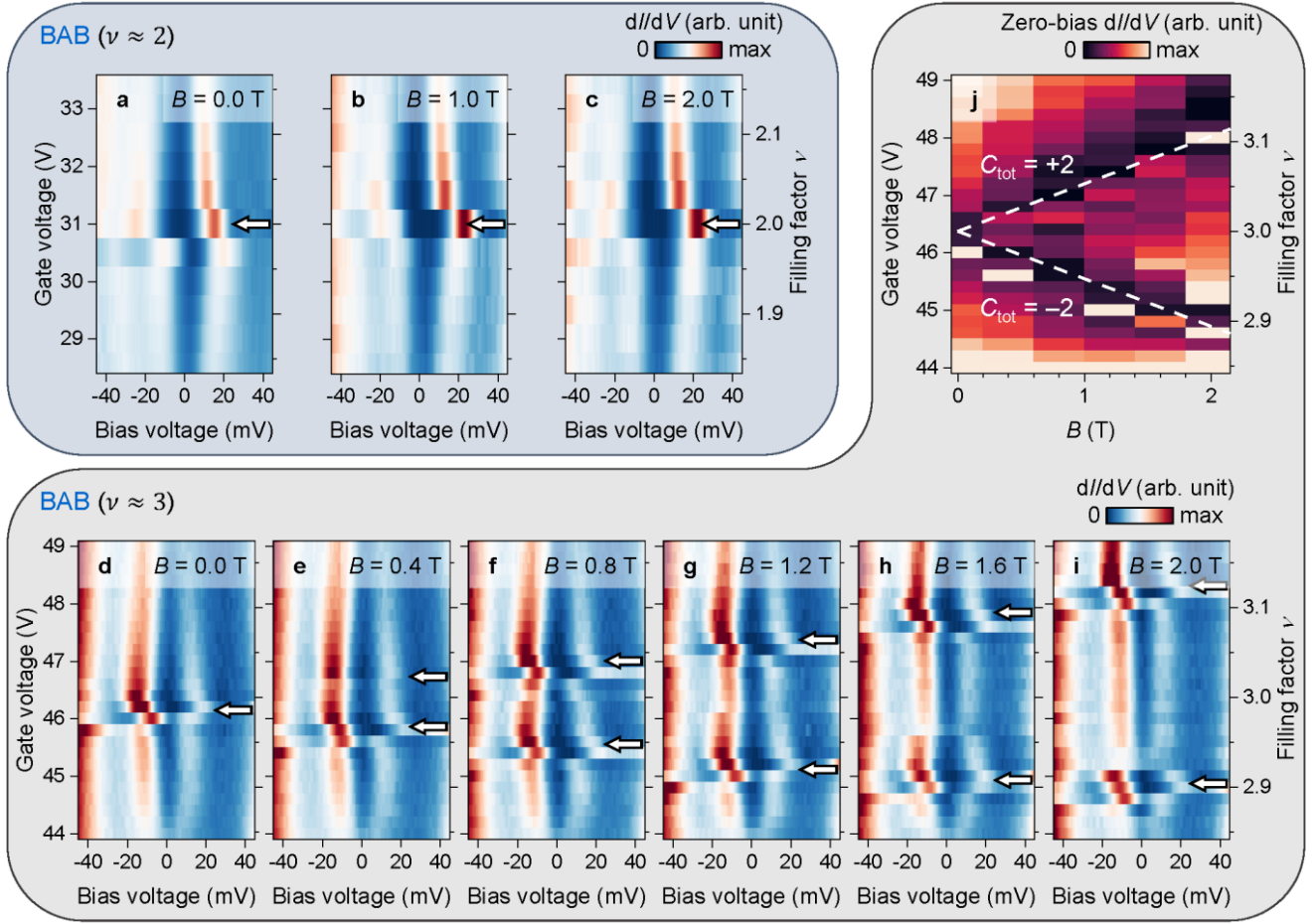

**Figure 2: Topological behaviour of correlated insulating states in an out-of-plane magnetic field. a-c**, Gate-dependent d$I$/d$V$ density plot for the BAB region near $v$ = 2 at (**a**) $B$ = 0.0 T, (**b**) $B$ = 1.0 T, and (**c**) $B$ = 2.0 T (modulation voltage $V_{RMS}$ = 1 mV; setpoint $V_{Bias}$ = –60 mV, $I_0$ = 0.5 nA). Arrows indicate correlation gaps. **d-i**, Gate-dependent d$I$/d$V$ density plot for the BAB region near $v$ = 3 at (**d**) $B$ = 0.0 T, (**e**) $B$ = 0.4 T, (**f**) $B$ = 0.8 T, (**g**) $B$ = 1.2 T, (**h**) $B$ = 1.6 T, and (**i**) $B$ = 2.0 T (modulation voltage $V_{RMS}$ = 1 mV; setpoint $V_{Bias}$ = –60 mV, $I_0$ = 0.1 nA). Arrows indicate correlation gaps. **j**, Normalized d$I$/d$V$ at $V_{Bias}$ = 0 mV ($E_F$) as a function of gate voltage and magnetic field. Dashed lines are guides to the eye following the Středa formula with total Chern number $C_{tot}$ = ±2.

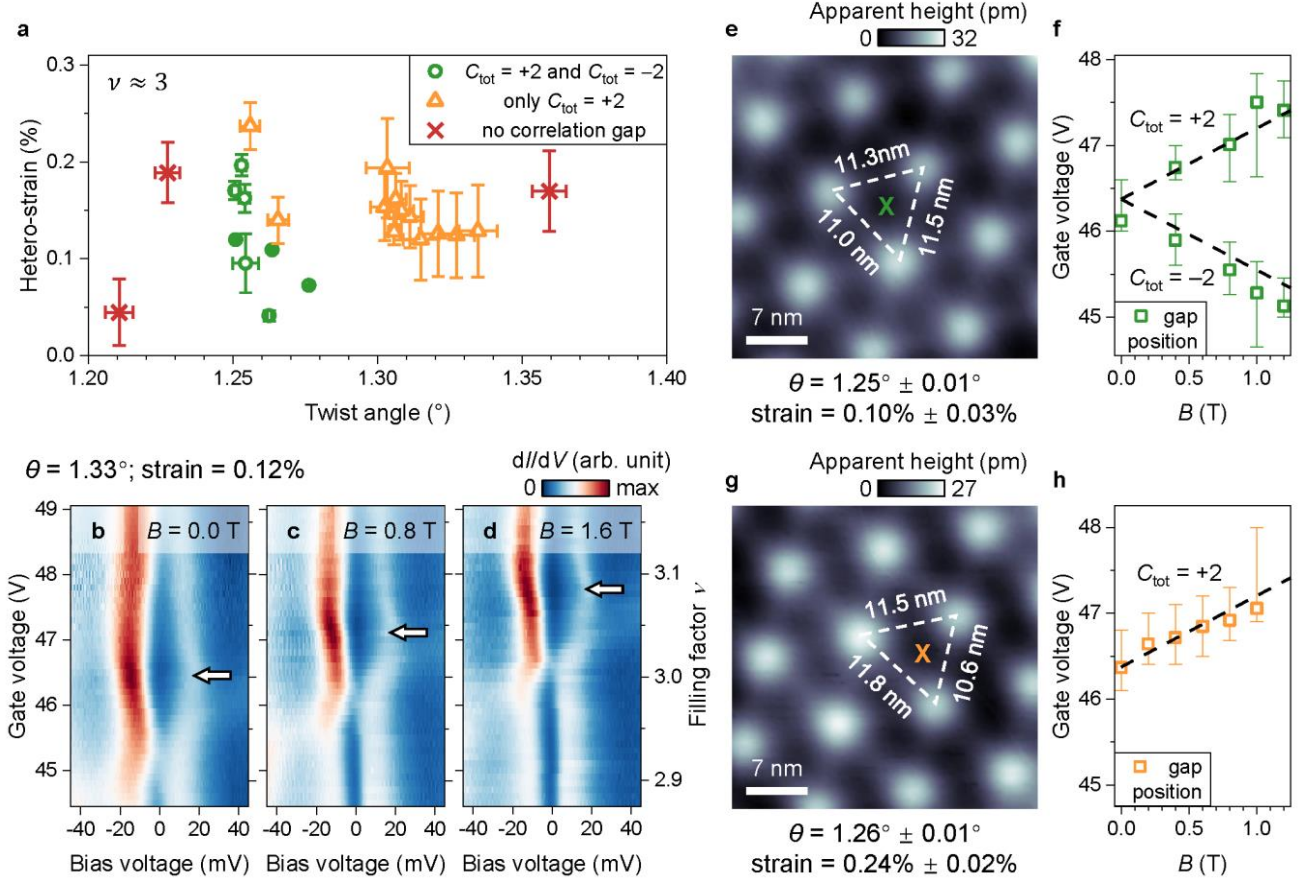

**Figure 3: Local structural effects on correlation and topology. a**, Electronic phase diagram of tMBLG at $v = 3$ in the parameter space of local twist angle and local hetero-strain. $C_{tot}$ is the total Chern number. **b-d**, Gate-dependent d$I$/d$V$ density plot for the BAB region at (**b**) $B = 0.0$ T, (**c**) $B = 0.8$ T, and (**d**) $B = 1.6$ T where only the $C_{tot} = +2$ gap is observed (modulation voltage $V_{RMS} = 1$ mV; setpoint $V_{Bias} = -75$ mV, $I_0 = 0.2$ nA). Arrows indicate correlation gaps. **e**, STM topographic image of a region with $\theta = 1.25°$ and small hetero-strain of 0.10% ($V_{Bias} = -1$ V, $I_0 = 0.02$ nA). Standard deviations for angle and strain are calculated from uncertainties in the moiré wavelength. **f**, Evolution of the correlation gap position near $v = 3$ in an out-of-plane magnetic field shows two branches for small strain. The data points were extracted from d$I$/d$V$ spectra taken in the BAB region marked in **e** with the error bars determined via linear fitting (Supplementary Note 5). Dashed lines are guides to the eye following the Středa formula with $C_{tot} = \pm 2$. **g**, STM topographic image of a region with $\theta = 1.26°$ and large hetero-strain of 0.24% ($V_{Bias} = -1$ V, $I_0 = 0.02$ nA). **h**, Evolution of the correlation gap position near $v = 3$ in an out-of-plane magnetic field shows only one branch for large strain. The data points were extracted from d$I$/d$V$ spectra taken in the BAB region marked in **g**. The dashed line is a guide to the eye following the Středa formula with $C_{tot} = +2$.

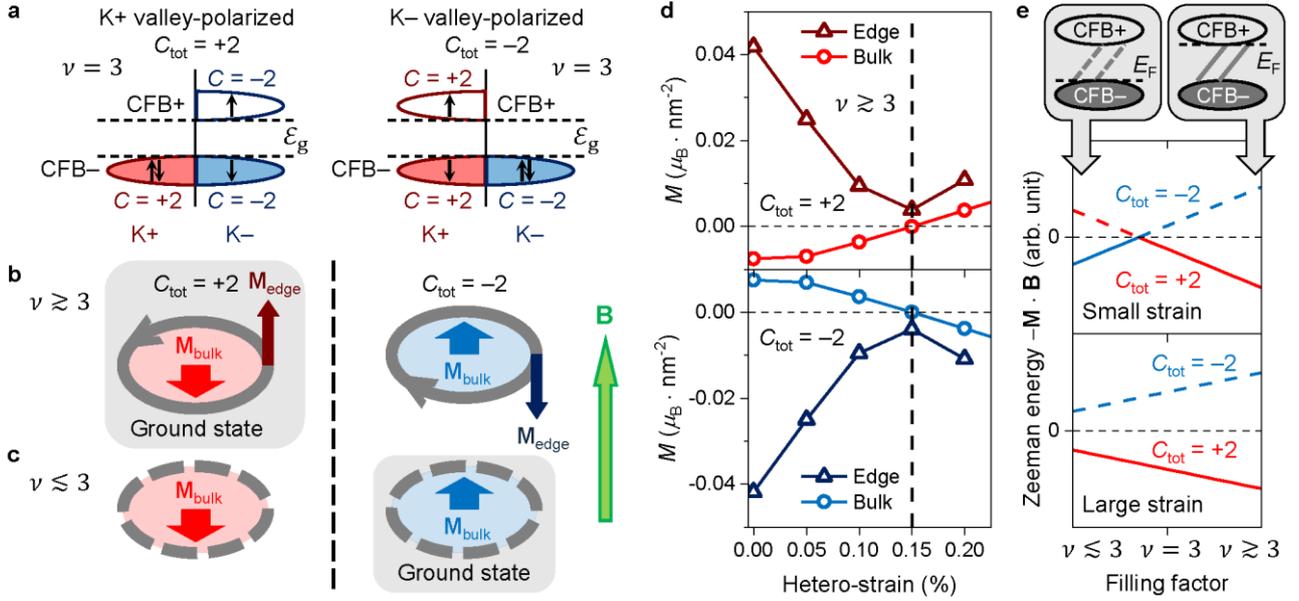

**Figure 4: Bulk-edge magnetization competition and gate-controlled Chern number switching**. **a**, Energy configuration of the K+ valley-polarized state and the K– valley-polarized state at $v = 3$ in tMBLG. Arrows represent electron spin. $\mathcal{E}_g$ is the size of the correlation gap. The total Chern number $C_{tot}$ is the sum of Chern numbers $C$ for all occupied sub-bands. **b**, Schematic of orbital magnetization for both valley-polarized states when $E_F$ is at the top of the correlation gap and the chiral edge state band is occupied ($v \gtrsim 3$). Grey arrows represent the direction of current flow in edge states. $\mathbf{M}_{bulk}$ and $\mathbf{M}_{edge}$ are bulk and edge orbital magnetic moments. Under an applied out-of-plane magnetic field **B** (green arrow on right) the $C_{tot} = +2$ state is the energetically favourable ground state. **c**, Same as **b**, but $E_F$ is at the bottom of the correlation gap and the edge state band is depleted ($v \lesssim 3$). Dashed grey lines represent empty edge states. Here the $C_{tot} = -2$ state is the ground state. **d**, Bulk and edge orbital magnetization for both valley-polarized states ($C_{tot} = \pm 2$) calculated based on a continuum model of 1.26° tMBLG. The vertical dashed line divides the parameter space into a small-strain regime where $M_{bulk}$ and $M_{edge}$ have opposite sign and a large-strain regime where they have the same sign. **e**, Top sketch shows the depletion/occupation of the edge states (represented by dashed/solid grey lines) for different filling levels. Orbital Zeeman energy is plotted for both valley-polarized states ($C_{tot} = \pm 2$) as a function of filling factor when $E_F$ is inside the correlation gap. Solid (dashed) lines represent the energetically favourable (unfavourable) states. CFB– = lower branch of conduction flat band, CFB+ = upper branch of conduction flat band.

**Supplementary Information: Local spectroscopy of a gate-switchable moiré quantum anomalous Hall insulator**

Canxun Zhang, Tiancong Zhu, Tomohiro Soejima, Salman Kahn, Kenji Watanabe, Takashi Taniguchi, Alex Zettl, Feng Wang, Michael P. Zaletel, Michael F. Crommie

**Table of Contents**

**Supplementary Note 1: Identifying the three stacking regions of tMBLG**

STM topographic images of tMBLG (Fig. 1b, Fig. 3e,g) typically show three distinct regions within each moiré unit cell, which we call "bright", "intermediate" and "dark" based on their apparent heights for $|V_{\text{Bias}}| > 100$ mV. We determined their local stacking orders by analysing structural as well as electronic contributions to the apparent height. AAB stacking in tMBLG is energetically unfavourable (due to strong repulsion between the carbon atoms in the "AA" layers) and exhibits an out-of-plane structural displacement, so we identify the "bright" region as AAB stacking. ABC and BAB stackings, on the other hand, have similar binding energies to each other and similar structural heights, so atomic structure alone is inadequate to explain the difference between the observed "intermediate" and "dark" regions. We have calculated the local density of states (LDOS) at the ABC and BAB sites using the continuum model (see Supplementary Note 6) and integrated them from –200 to 0 meV (simulating a negative $V_{\text{Bias}}$) or from 0 to 200 meV (simulating a positive $V_{\text{Bias}}$), and find that ABC always displays a higher intensity than BAB. Since a larger integrated LDOS corresponds to a higher apparent height in constant-current STM measurements, we identify the "dark" region as corresponding to BAB stacking and the "intermediate" region ABC stacking.

**Supplementary Note 2: Comparison between transport and STM/STS measurements**

In transport measurements on tMBLG the graphene layers are usually encapsulated between two pieces of hBN with a top gate and a bottom gate on two sides.[1-4] This allows independent tuning of carrier density $n$ and out-of-plane electric field $\mathbf{E} = (0, 0, E)$ in the graphene stack through combination of top and bottom gate voltages. The electric field creates a potential difference $\Delta_U = eEd_a$ between adjacent graphene layers ($d_a = 0.33$ nm is the inter-layer distance) which impacts the shape and alignment of flat bands and hence the correlated states of tMBLG. Supplementary Figure 2 shows a schematic of $R_{xx}$ as a function of both $n$ and $E$ in a typical transport measurement of tMBLG (from Ref. [1]). Correlated insulating states emerge at $v = 1, 2, 3$ over a finite range of $E$, demonstrating electric-field-tuning of correlation effects in tMBLG. In our STM/STS measurement geometry, a dedicated top gate is not achievable due to the existence of the STM tip. Both $n$ and $E$ are controlled by the back-gate voltage $V_G$ and cannot be independently tuned (tip-induced gating is not observed in our experiment due to deliberate work function matching between graphene and the tip material). The different chemical environment of exposed carbon atoms in the top layer and those in contact with hBN in the bottom layer leads to an additional inter-layer potential difference $\Delta_{U0}$. The overall inter-layer potential difference is $\Delta_U = \Delta_{U0} + eEd_a$ where $E$ is directly related to $n$ by $E = \frac{ne}{2\varepsilon_{\text{eff}}\varepsilon_0}$ ($\varepsilon_{\text{eff}}$ is the effective out-of-plane dielectric constant of tMBLG). The parameter space in STM/STS measurements therefore corresponds to a "diagonal" line-cut in Supplementary Fig. 2, which touches the $v = 2$ and the $v = 3$ correlated insulating states while missing the $v = 1$ one. This explains why correlation gaps appear in d$I$/d$V$ at $v = 2, 3$ but not at $v = 1$ in our data.

**Supplementary Note 3: Response of the $v = 3$ correlation gap size to an applied out-of-plane magnetic field**

The size of the $v = 3$ correlation gap should be modified by application of an out-of-plane magnetic field $\mathbf{B} = (0, 0, B)$ due to the Zeeman energy associated with non-zero orbital magnetic moments, analogous to the case in Ref. [5]. We can estimate the magnitude of this effect by considering single-particle effects and assuming that each $C = +2$ sub-band in the CFB manifold has a magnetization of $m_{C = \pm 2}$ as defined in Supplementary Note 6. For the K+ valley-polarized $C_{\text{tot}} = +2$ state (Fig. 4a), the doubly-occupied $C = +2$ sub-bands then shift downward by $B|m_{C = \pm 2}|$ (assuming $B > 0$), whereas both the occupied and the unoccupied $C = -2$ sub-bands shift upward by the same amount. The correlation gap should remain the same since it lies between the two $C = -2$ sub-bands. For the K– valley-polarized $C_{\text{tot}} = -2$ state (Fig. 4b), on the other hand, the correlation gap (which lies between the doubly occupied $C = -2$ sub-bands and the unoccupied $C = +2$ sub-band) should decrease by $2B|m_{C = \pm 2}|$. We estimate this shift to be less than 2 meV at $B = 2$ T based on calculated $m_{C = \pm 2}$ values (Supplementary Fig. 9b-d), and so this effect is obscured by thermal and instrumental broadening and is not significant in our data.

**Supplementary Note 4: Comparison between QAH phases in tMBLG and MA-tBLG**

In addition to tMBLG, the QAH effect has also been observed in magic-angle twisted bilayer graphene (MA-tBLG) aligned with an hBN substrate.[6] The underlying physics of the QAH phases in tMBLG and hBN-aligned MA-tBLG are similar, but two major differences have been reported in transport experiments. First, the Hall conductance is quantized to $\pm e^2/h$ for MA-tBLG and $\pm 2e^2/h$ for tMBLG, since the Chern number for QAH states of MA-tBLG is $\pm 1$ while for tMBLG it is $\pm 2$. Second, gate-induced Chern number switching has only been observed in tMBLG. This likely depends on the details of band structure, as discussed in Ref. [7]. In terms of local spectroscopy, signatures of Chern insulating behaviour have been detected via STM/STS in hBN-aligned MA-tBLG under low magnetic fields,[8] but no systematic study on the QAH phase has been performed yet.

**Supplementary Note 5: Determining the correlation gap position**

We determined the $V_G$ ($\nu$) values at which the correlation gap appears (Fig. 3f,h) using the following procedure. Supplementary Figure 7a shows normalized d$I$/d$V$ at $V_{Bias} = 0$ mV ($E_F$) as a function of $V_G$ for different $B$ values (this is for a small-strain region; the data is the same as in Fig. 2j). The single dip at $\nu = 3$ for $B = 0.0$ T and two separate dips above and below $\nu = 3$ for finite $B$ signify the formation of correlation gaps at these filling factors. The dip features, however, are often accompanied by prominent peaks indicated by red arrows in Supplementary Fig. 7a. To understand their origin, we carefully examine Fig. 2d-i (Fig. 2d is reproduced with slightly different color scale in Supplementary Fig. 7b as an example) and focus on the features that give rise to the dip-adjacent peaks in Supplementary Fig. 7a. As $V_G$ increases, they rapidly shift to higher energies (highlighted by the dashed box in Supplementary Fig. 7b), which is the opposite direction compared to the movement of CFB and VFB peaks. This suggests that they likely arise from tip-induced charging[9,10] and do not reflect the actual LDOS in tMBLG. For every curve in Supplementary Fig. 7a, we thus exclude the dip-adjacent peaks and perform linear fitting on the two sides of each dip (fitted lines are shown for $B = 0.0$ T as an example). The data points in Fig. 3f indicate intersection of fitted lines and the error bars come from the fitting errors. The same procedure is conducted for data from the large-strain region (Supplementary Fig. 7c; the data is the same as in Supplementary Fig. 6f) and the results are plotted in Fig. 3h.

**Supplementary Note 6: Theoretical model and calculations**

**Continuum model of tMBLG.** Our calculations were based on the Bistritzer–MacDonald continuum approach to moiré structures.[11,12] Here the monolayer graphene is modelled using a two-band tight-binding model with $t_0 = 2.8$ eV while the Bernal-stacked bilayer is modelled using a four-band model with $t_0 = 2.61$ eV, $t_1 = 0.361$ eV, $t_3 = 0.283$ eV, $t_4 = 0.138$ eV, $\Delta = 0.015$ eV.[13] The bilayer is then rotated by angle $\theta$ and hybridized with the monolayer with intra-sublattice strength $w_{AA} = 87.75$ meV and inter-sublattice strength $w_{AB} = 117$ meV. The hetero-strain is taken into consideration as an artificial vector field, following the treatment in Ref. [14]. We further add a potential difference $\Delta_U = (12.6 + 4.69\,\nu)$ meV between adjacent graphene layers to account for the gate-induced out-of-plane electric field as well as the built-in asymmetry between top and bottom layers due to the presence of the hBN substrate (Supplementary Note 2). The resulting continuum model is truncated by keeping all states within a radius of 6 mini-Brillouin zones (mBZs) of the $\gamma$-point.

**Decomposing orbital magnetization into bulk and edge components.** When the chemical potential $\mu$ resides inside the $\nu = 3$ correlation gap, the total orbital magnetization can be decomposed into the bulk part which is independent of $\mu$ and the edge part which depends linearly on $\mu$. There is no unique way of defining these two pieces. We find it convenient to use the following convention:

$$M_{\text{bulk}} = \frac{e}{2\hbar} \sum_n \int d^2\mathbf{k} \, \epsilon^{ab} \, \text{Im} \langle \partial_a u_{n,\mathbf{k}} | H_{n,\mathbf{k}} + \mathcal{E}_{n,\mathbf{k}} - 2\mathcal{E}_{\text{max}}^{\text{OC}} | \partial_b u_{n,\mathbf{k}} \rangle \tag{1}$$

$$M_{\text{edge}}(\mu) = -\frac{e}{\hbar} \sum_n \int d^2\mathbf{k} \, \epsilon^{ab} \, \text{Im} \langle \partial_a u_{n,\mathbf{k}} | \mu - \mathcal{E}_{\text{max}}^{\text{OC}} | \partial_b u_{n,\mathbf{k}} \rangle = \frac{e}{h} C_{\text{tot}} (\mu - \mathcal{E}_{\text{max}}^{\text{OC}}) \tag{2}$$

where $\epsilon^{ab}$ is the antisymmetric tensor, $\mathcal{E}_{n,\mathbf{k}}$ is the single particle energy, and $\mathcal{E}_{\text{max}}^{\text{OC}}$ is the energy of the top of the occupied bands (i.e., the bottom of the correlation gap). This convention has the advantage that $M_{\text{edge}}(\mathcal{E}_{\text{max}}^{\text{OC}}) = 0$, as befitting the interpretation that $M_{\text{edge}}$ is the contribution from chiral edge modes. Appropriate care needs to be taken to compute these quantities on a discrete grid, as described earlier in Refs. [15,16].

**Calculating $M_{\text{edge}}$.** For a given band structure, the sign of $M_{\text{edge}}$ always follows that of $C_{\text{tot}}$, while its magnitude reaches the maximum when $\mu = \mathcal{E}_{\text{min}}^{\text{UO}}$ ($\mathcal{E}_{\text{min}}^{\text{UO}}$ is the energy of the bottom of the unoccupied bands, i.e., the top of the correlation gap):

$$M_{\text{edge}}(\mu = \mathcal{E}_{\text{min}}^{\text{UO}}) = \frac{e}{h} C_{\text{tot}} (\mathcal{E}_{\text{min}}^{\text{UO}} - \mathcal{E}_{\text{max}}^{\text{OC}}) = \frac{e}{h} C_{\text{tot}} \mathcal{E}_g \tag{3}$$

where $\mathcal{E}_g$ is the size of the correlation gap. To obtain $\mathcal{E}_g$ values as a function of hetero-strain, we performed Hartree-Fock calculation in a momentum-space approach analogous to earlier Hartree-Fock studies of twisted graphene systems. Our code is an extension of the tBLG code used in Ref. [17], the twisted double bilayer graphene (tDBLG) code used in Ref. [18], and the tMBLG code used in Ref. [19]. The Coulomb interaction (screened by the graphene, the hBN/SiO$_2$ substrate and the metallic STM tip) is assumed to take the single-plane-screened form $V(\mathbf{q}) = \frac{e^2}{2\epsilon_{\text{eff}}\epsilon_0 q}[1 - \exp(-2qd_S)]$ ($\epsilon_{\text{eff}}$ = 12 and $d_S$ = 20 nm were chosen on phenomenological grounds). The Coulomb matrix elements are evaluated in the basis of the continuum band structure and projected into the six bands nearest to the charge neutrality per valley and spin, for a total of 24 bands. We then consider a Slater-determinant ansatz $|u\rangle$ that is diagonal in the mBZ momentum $\mathbf{k}$. Discretizing the model on a 16x16 $\mathbf{k}$-grid, the $u$'s are iteratively adjusted to minimize the energy $\langle u|H|u\rangle$, using the optimal damping algorithm in Ref. [20] to achieve Hartree-Fock self-consistency. Supplementary Figure 8a-c shows the resulting Hartree-Fock tMBLG band structures for selected hetero-strain values. Here positive (negative) strain is defined as stretching (compressing) the bilayer and compressing (stretching) the monolayer (bilayer). Increasing the hetero-strain leads to a reduction in $\mathcal{E}_g$ (Supplementary Fig. 8d) and also in $M_\text{edge}$ as plotted in Fig. 4d (positive strain is assumed in the main text, but the results remain qualitatively the same for negative strain).

**Calculating $M_\text{bulk}$.** As $M_\text{bulk}$ is heavily dependent on the detailed band structure, a full calculation based on the Hartree-Fock approach is extremely costly and beyond our current capabilities. Instead, we adopt the phenomenological model proposed in Refs. [1,7] to take electron-electron interaction into consideration. For our purpose it is convenient to take an alternative decomposition of orbital magnetization

$$M_\text{bulk} = \frac{e}{2\hbar} \sum_n \int d^2\mathbf{k}\, \epsilon^{ab}\, \text{Im}\langle \partial_a u_{n,\mathbf{k}} | H_{n,\mathbf{k}} + \mathcal{E}_{n,\mathbf{k}} - 2\mathcal{E}_{\max,n} | \partial_b u_{n,\mathbf{k}} \rangle$$
$$- \frac{e}{\hbar} \sum_n \int d^2\mathbf{k}\, \epsilon^{ab}\, \text{Im}\langle \partial_a u_{n,\mathbf{k}} | \mu - \mathcal{E}_{\max,n} | \partial_b u_{n,\mathbf{k}} \rangle \quad (4)$$
$$= \sum_n m_n + \frac{e}{h} \sum_n C_n (\mu - \mathcal{E}_{\max,n})$$

where $\mathcal{E}_{\max,n}$ is the energy of the top of the $n$-th band and $m_n$ can be interpreted as the total orbital magnetization of that band when the chemical potential is at $\mathcal{E}_{\max,n}$. We start from the single-particle band structure (for which the $m_n$'s can be straightforwardly calculated using the dual state method of Ref. [16]) and analyse how electron-electron interaction impacts $M_\text{bulk}$. The primary effect of interaction is to introduce exchange energy difference between different sub-bands which causes spontaneous polarization in the spin-valley space. We expect the nature of each sub-band to remain unchanged to a good approximation. Therefore, we make several simplifying assumptions similar to those in Ref. [1] with slight modifications:

1. Interaction opens up a gap between the three occupied CFB− sub-bands and the one unoccupied CFB+ sub-band;
2. Interaction introduces an exchange-driven energy difference $\delta_\text{VFB}$ among the VFB sub-bands (which have valley-dependent $C = \mp 1$) (Supplementary Fig. 9a);
3. The energies of other bands remain unchanged;
4. Each $m_n$, evaluated at the top of each band after taking $\delta_\text{VFB}$ into account, remains unchanged due to interaction.

We can evaluate the total orbital magnetization when $\mu = \mathcal{E}_\text{max}^\text{OC}$ with these assumptions. Focusing on the K+ valley-polarized $C_\text{tot} = +2$ state, we get

$$\begin{aligned}
M_{\text{bulk}} &= \sum_n m_n + \frac{e}{h}\sum_n C_n\big(\mathcal{E}_{\max}^{\text{OC}} - \mathcal{E}_{\max,n}\big) \\
&= m_{C=+2} + m_{C=+2} + m_{C=-2} \\
&\quad + \frac{e}{h}\big[(-1)\big(\mathcal{E}_{\max}^{\text{OC}} - \mathcal{E}_{\max,\text{VFB}}\big) + (-1)\big(\mathcal{E}_{\max}^{\text{OC}} - \mathcal{E}_{\max,\text{VFB}}\big) \\
&\quad + (1)\big(\mathcal{E}_{\max}^{\text{OC}} - \mathcal{E}_{\max,\text{VFB}}\big) + (1)\big(\mathcal{E}_{\max}^{\text{OC}} - \mathcal{E}_{\max,\text{VFB}} - \delta_{\text{VFB}}\big)\big] \\
&= m_{C=+2} - \frac{e}{h}\delta_{\text{VFB}}
\end{aligned} \quad (5)$$

$M_{\text{bulk}}$ for the $C_{\text{tot}} = -2$ state is the exact opposite. Here the first term (arising from single-particle effects) is parallel to $M_{\text{edge}}$ while the second term characterizing electron-electron interaction strength is antiparallel to $M_{\text{edge}}$.

We find that $m_{C=+2}$ becomes larger as the hetero-strain is increased due to a redistribution of Berry curvature and band dispersion throughout the mBZ. $\delta_{\text{VFB}}$ is treated as a fitting parameter since it is difficult to accurately estimate the exchange coupling. Supplementary Figure 9b-d shows $M_{\text{bulk}}(C_{\text{tot}} = +2)$ as a function of the hetero-strain for different choices of $\delta_{\text{VFB}}$ when the twist angle $\theta$ is fixed at 1.26°, 1.28°, and 1.32°, respectively. In the main text we use $\delta_{\text{VFB}} = 13$ meV for $\theta = 1.26°$ (Fig. 4d), which best matches our observation that increasing hetero-strain causes $M_{\text{bulk}}(C_{\text{tot}} = +2)$ to flip from negative to positive and the system to change from switchable to non-switchable. For $1.30° < \theta < 1.34°$, only non-switchable QAH states have been observed experimentally (Fig. 3a), but this does not preclude a similar strain-induced transition since $M_{\text{bulk}}$ could flip its sign for strains below 0.12%, the lowest amount observed within this angle range.

**Supplementary References**

1. Polshyn, H. *et al.* Electrical switching of magnetic order in an orbital Chern insulator. *Nature* **588**, 66-70 (2020). https://doi.org:10.1038/s41586-020-2963-8

2. Chen, S. *et al.* Electrically tunable correlated and topological states in twisted monolayer–bilayer graphene. *Nature Physics* **17**, 374-380 (2021). https://doi.org:10.1038/s41567-020-01062-6

3. Xu, S. *et al.* Tunable van Hove singularities and correlated states in twisted monolayer–bilayer graphene. *Nature Physics* **17**, 619-626 (2021). https://doi.org:10.1038/s41567-021-01172-9

4. He, M. *et al.* Competing correlated states and abundant orbital magnetism in twisted monolayer-bilayer graphene. *Nature Communications* **12**, 4727 (2021). https://doi.org:10.1038/s41467-021-25044-1

5. Li, S.-Y. *et al.* Experimental evidence for orbital magnetic moments generated by moiré-scale current loops in twisted bilayer graphene. *Physical Review B* **102**, 121406 (2020). https://doi.org:10.1103/PhysRevB.102.121406

6    Serlin, M. *et al.* Intrinsic quantized anomalous Hall effect in a moiré heterostructure. *Science* **367**, 900-903 (2020). https://doi.org:10.1126/science.aay5533

7    Zhu, J., Su, J.-J. & MacDonald, A. H. Voltage-Controlled Magnetic Reversal in Orbital Chern Insulators. *Physical Review Letters* **125**, 227702 (2020). https://doi.org:10.1103/PhysRevLett.125.227702

8    Oh, M. *et al.* Evidence for unconventional superconductivity in twisted bilayer graphene. *Nature* **600**, 240-245 (2021). https://doi.org:10.1038/s41586-021-04121-x

9    Zhao, Y. *et al.* Creating and probing electron whispering-gallery modes in graphene. *Science* **348**, 672-675 (2015). https://doi.org:10.1126/science.aaa7469

10   Velasco, J., Jr. *et al.* Visualization and Control of Single-Electron Charging in Bilayer Graphene Quantum Dots. *Nano Letters* **18**, 5104-5110 (2018). https://doi.org:10.1021/acs.nanolett.8b01972

11   Park, Y., Chittari, B. L. & Jung, J. Gate-tunable topological flat bands in twisted monolayer-bilayer graphene. *Physical Review B* **102**, 035411 (2020). https://doi.org:10.1103/PhysRevB.102.035411

12   Rademaker, L., Protopopov, I. V. & Abanin, D. A. Topological flat bands and correlated states in twisted monolayer-bilayer graphene. *Physical Review Research* **2**, 033150 (2020). https://doi.org:10.1103/PhysRevResearch.2.033150

13   Jung, J. & MacDonald, A. H. Accurate tight-binding models for the π bands of bilayer graphene. *Physical Review B* **89**, 035405 (2014). https://doi.org:10.1103/PhysRevB.89.035405

14   Bi, Z., Yuan, N. F. Q. & Fu, L. Designing flat bands by strain. *Physical Review B* **100**, 035448 (2019). https://doi.org:10.1103/PhysRevB.100.035448

15   Souza, I., Íñiguez, J. & Vanderbilt, D. Dynamics of Berry-phase polarization in time-dependent electric fields. *Physical Review B* **69**, 085106 (2004). https://doi.org:10.1103/PhysRevB.69.085106

16   Ceresoli, D., Thonhauser, T., Vanderbilt, D. & Resta, R. Orbital magnetization in crystalline solids: Multi-band insulators, Chern insulators, and metals. *Physical Review B* **74**, 024408 (2006). https://doi.org:10.1103/PhysRevB.74.024408

17   Bultinck, N. *et al.* Ground State and Hidden Symmetry of Magic-Angle Graphene at Even Integer Filling. *Physical Review X* **10**, 031034 (2020). https://doi.org:10.1103/PhysRevX.10.031034

18   Zhang, C. *et al.* Visualizing delocalized correlated electronic states in twisted double bilayer graphene. *Nature Communications* **12**, 2516 (2021). https://doi.org:10.1038/s41467-021-22711-1

19   Polshyn, H. *et al.* Topological charge density waves at half-integer filling of a moiré superlattice. *Nature Physics* **18**, 42-47 (2022). https://doi.org:10.1038/s41567-021-01418-6

20   Cancès, E. & Le Bris, C. Can we outperform the DIIS approach for electronic structure calculations? *International Journal of Quantum Chemistry* **79**, 82-90 (2000). https://doi.org:https://doi.org/10.1002/1097-461X(2000)79:2<82::AID-QUA3>3.0.CO;2-I

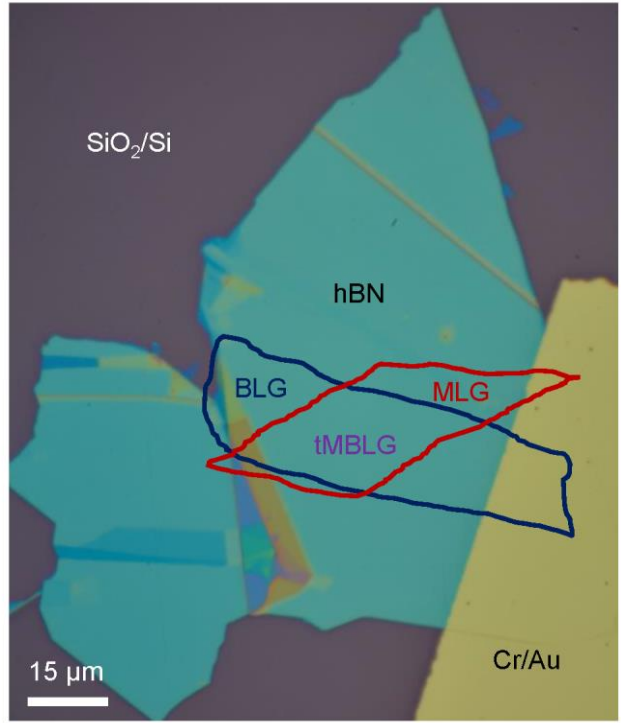

**Supplementary Figure 1: Optical microscope image of the tMBLG device.** MLG = monolayer graphene; BLG = Bernal-stacked bilayer graphene.

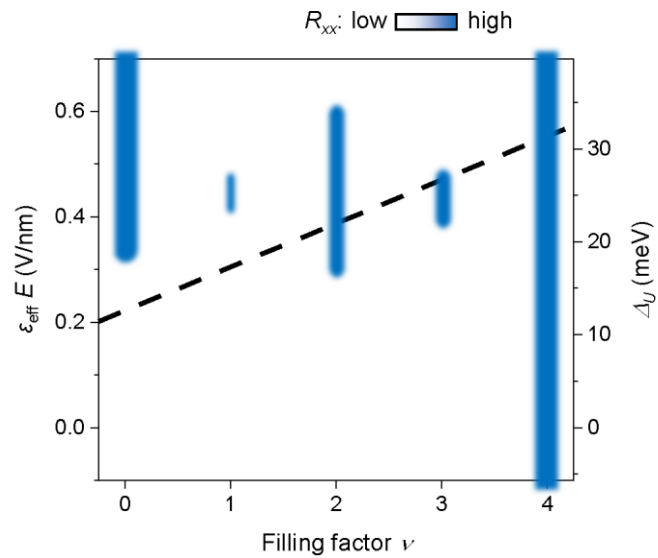

**Supplementary Figure 2: Parameter space in transport and STM/STS measurements.** The dark blue regions indicate insulating phases in transport measurement (adapted from Ref. [1]). The dashed line is the STM/STS parameter space. $\varepsilon_{eff}$ is the effective dielectric constant of tMBLG, $E$ is the out-of-plane electric field, and $\Delta_U$ is the corresponding inter-layer potential difference.

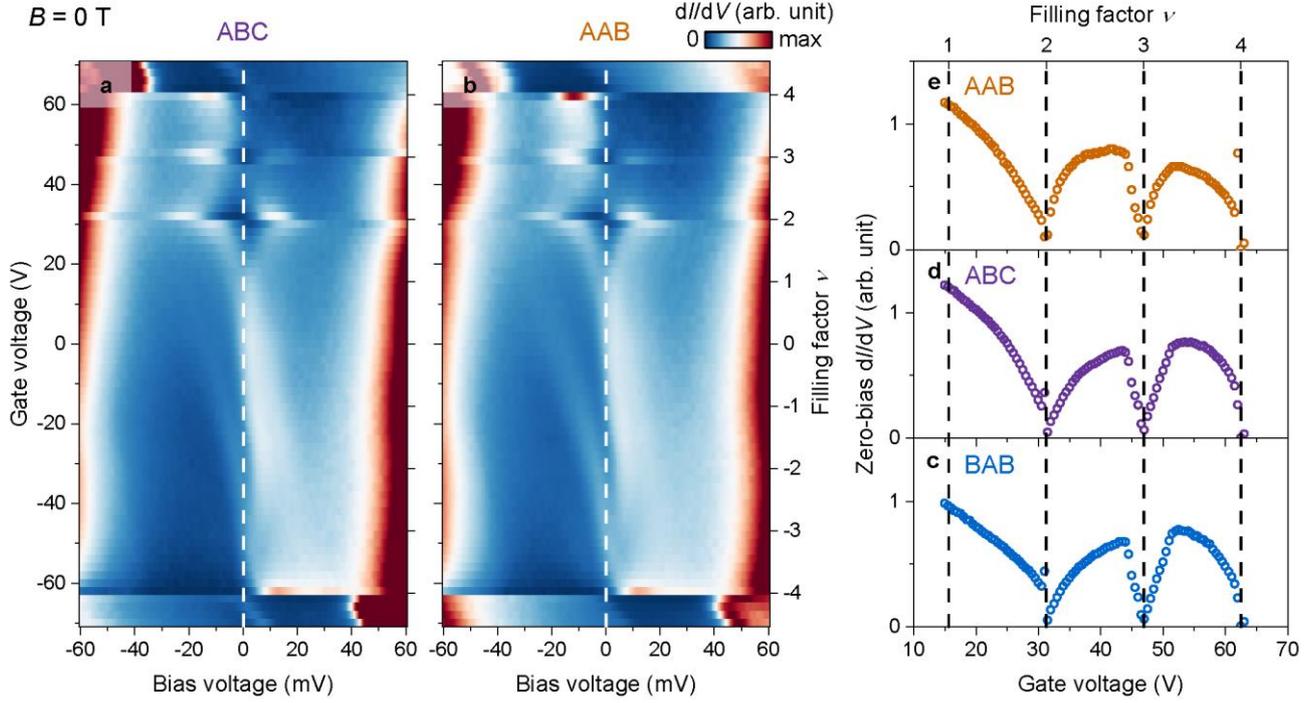

**Supplementary Figure 3: Additional data for correlated insulating states at $v = 2, 3$. a**, Gate-dependent $dI/dV$ density plot for the ABC stacking region over the gate range $-70 \text{ V} \leq V_G \leq 70 \text{ V}$. The vertical dashed line denotes the Fermi energy $E_F$. Spectroscopy parameters: modulation voltage $V_{RMS} = 1$ mV; setpoint $V_{Bias} = 100$ mV, $I_0 = 1.55$ nA for $-70 \text{ V} \leq V_G \leq -2$ V; setpoint $V_{Bias} = -100$ mV, $I_0 = 0.8$ nA for $0 \text{ V} \leq V_G \leq 70$ V. **b**, Same as **a**, but for the AAB stacking region. Spectroscopy parameters: modulation voltage $V_{RMS} = 1$ mV; setpoint $V_{Bias} = 100$ mV, $I_0 = 1.4$ nA for $-70 \text{ V} \leq V_G \leq -2$ V; setpoint $V_{Bias} = -100$ mV, $I_0 = 0.8$ nA for $0 \text{ V} \leq V_G \leq 70$ V. Splitting of the CFB peak around $v = 2, 3$ is observed in spectra measured in all three regions (see also Fig. 1c). **c-e**, Normalized $dI/dV$ at $V_{Bias} = 0$ mV ($E_F$) as a function of $V_G$ ($v$) for all three stacking regions. This type of plot is analogous to a transport conductance measurement since in both cases only electronic states near $E_F$ are being probed. Dips going down to nearly vanishing $dI/dV$ that are centred at $v = 2, 3$, and 4 indicate emergence of insulating phases at these filling factors.

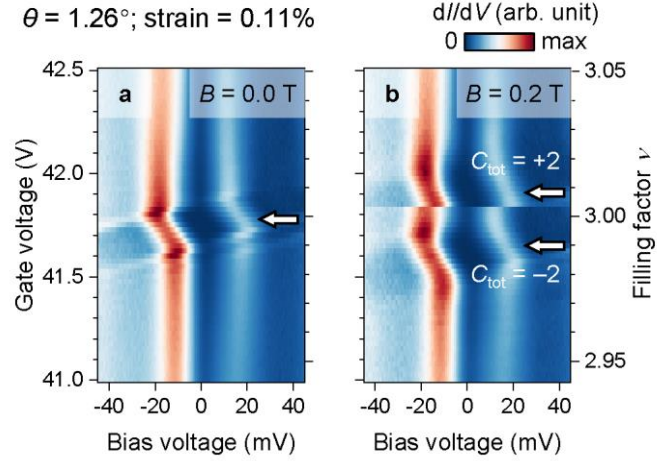

**Supplementary Figure 4: Topological behaviour of the $\nu$ = 3 state at $B$ = 0.2 T.** Gate-dependent d$I$/d$V$ density plot near $\nu$ = 3 at (**a**) $B$ = 0.0 T and (**b**) $B$ = 0.2 T ($V_{RMS}$ = 1 mV; setpoint $V_{Bias}$ = –60 mV, $I_0$ = 0.5 nA). Arrows indicate correlation gaps.

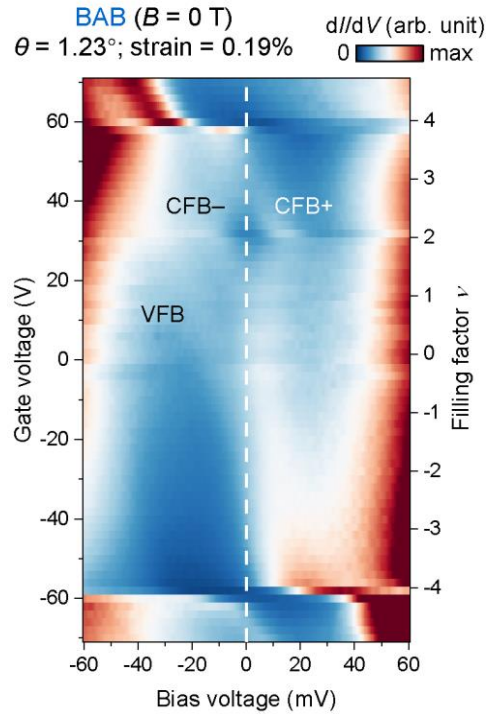

**Supplementary Figure 5: Gate-dependent d$I$/d$V$ without the $\nu$ = 3 correlation gap.** These data were obtained in a region with $\theta$ = 1.23° and hetero-strain = 0.19%. The vertical dashed line denotes the Fermi energy. Spectroscopy parameters: modulation voltage $V_{RMS}$ = 1 mV; setpoint $V_{Bias}$ = 100 mV, $I_0$ = 2 nA for –70 V ≤ $V_G$ ≤ –2 V; setpoint $V_{Bias}$ = –100 mV, $I_0$ = 1 nA for 0 V ≤ $V_G$ ≤ 70 V. VFB = valence flat band, CFB– = lower branch of conduction flat band, CFB+ = upper branch of conduction flat band.

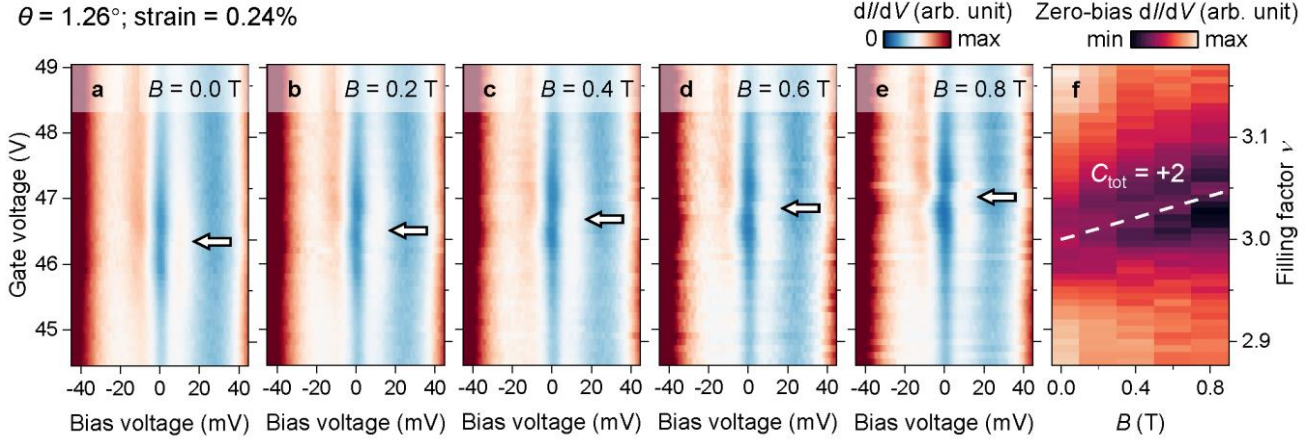

**Supplementary Figure 6: Topological behaviour of the $v = 3$ state for the large-strain region. a-e**, Gate-dependent d$I$/d$V$ density plot near $v = 3$ for a region with $\theta = 1.26°$ and hetero-strain = 0.24% at (**a**) $B = 0.0$ T, (**b**) $B = 0.2$ T, (**c**) $B = 0.4$ T, (**d**) $B = 0.6$ T, and (**e**) $B = 0.8$ T (modulation voltage $V_{RMS} = 1$ mV; setpoint $V_{Bias} = -60$ mV, $I_0 = 0.3$ nA). Arrows indicate correlation gaps. **f**, Normalized d$I$/d$V$ at $V_{Bias} = 0$ mV ($E_F$) as a function of $V_G$ ($v$) and $B$. The dashed line is a guide to the eye following the Středa formula with $C_{tot} = +2$.

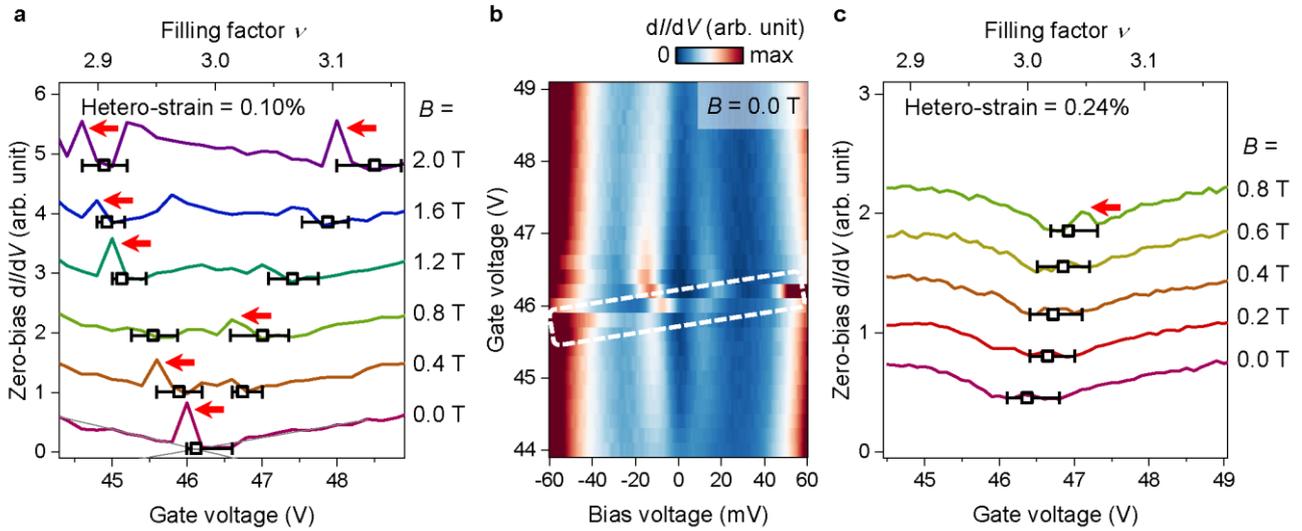

**Supplementary Figure 7: Extracting correlation gap positions from gate-dependent d$I$/d$V$. a**, Normalized d$I$/d$V$ at $V_{Bias} = 0$ mV ($E_F$) as a function of $V_G$ ($v$) for different magnetic fields (small-strain region). Red arrows indicate tip-induced charging features. Squares with error bars are correlation gap positions. Fitted lines are plotted in grey for $B = 0.0$ T. **b**, Reproduction of gate-dependent d$I$/d$V$ in Fig. 2d with the charging feature highlighted (dashed white box). **c**, same as **a**, but with data from the large-strain region.

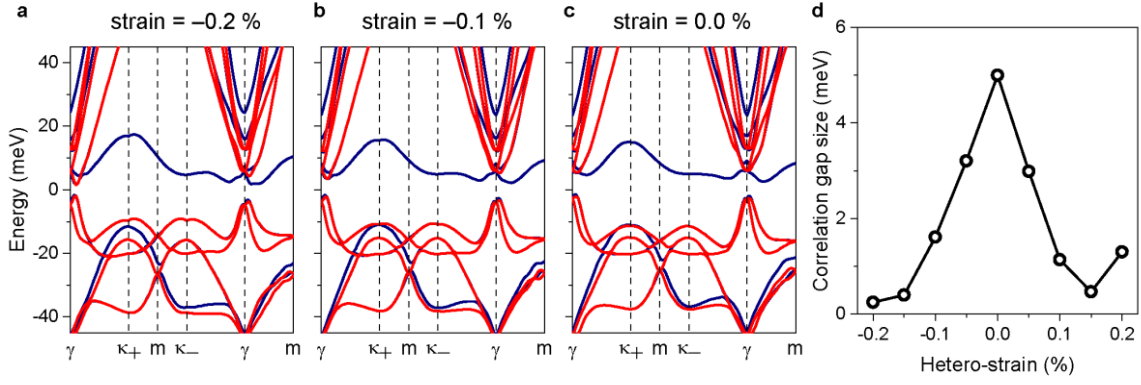

**Supplementary Figure 8: Strain tuning Hartree-Fock band structure for tMBLG. a-c**, Hartree-Fock band structures plotted along a high-symmetry line in the mBZ for hetero-strains of (**a**) –0.2%, (**b**) –0.1%, and (**c**) 0.0%. Red and blue curves represent sub-bands with different spins. **d**, Extracted correlation gap size as a function of hetero-strain.

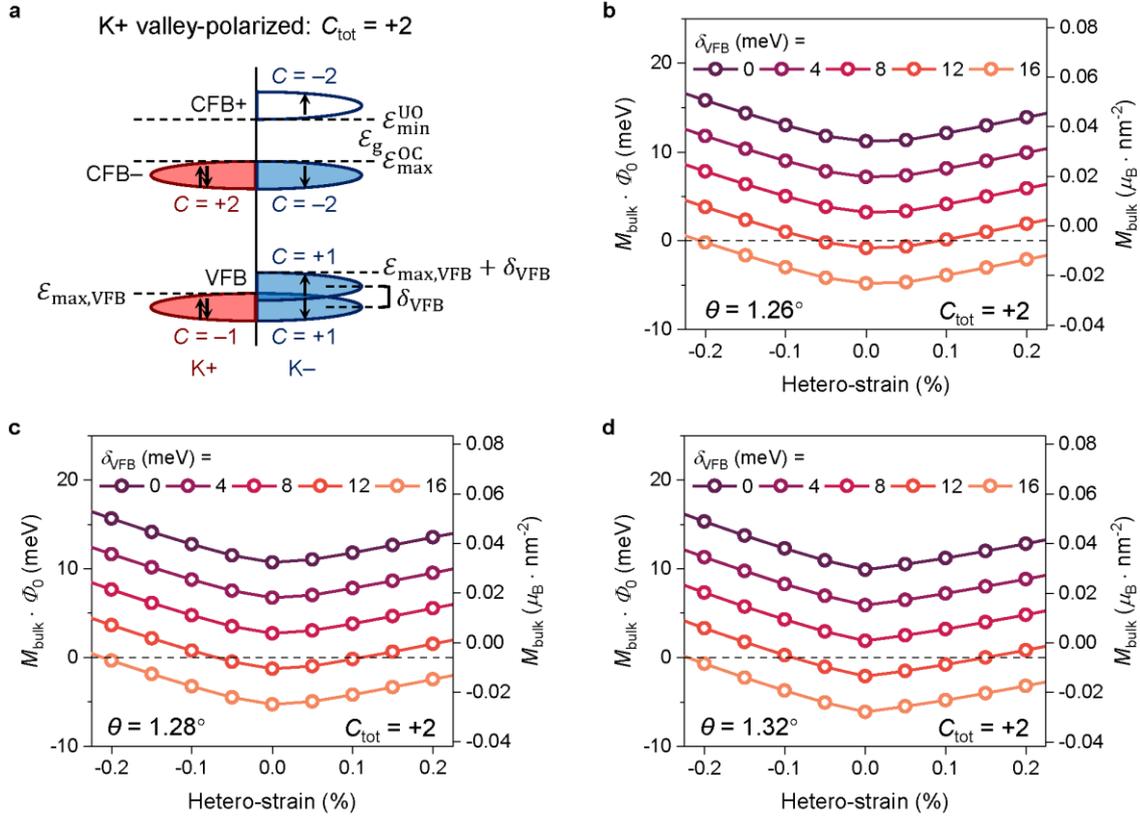

**Supplementary Figure 9: Calculating bulk magnetization in the presence of electron-electron interaction. a**, Energy configuration of the K+ valley-polarized state at $\nu = 3$ showing sub-bands from both the CFB manifold and the VFB manifold. $\delta_{VFB}$ is the interaction-induced energy offset among the VFB sub-bands. **b-d**, $M_{bulk}$ for $C_{tot} = +2$ state as a function of hetero-strain for different choices of $\delta_{VFB}$ at (**b**) $\theta = 1.26°$, (**c**) 1.28°, and (**d**) 1.32°.

13